\documentclass[aps,prd,twocolumn,floatfix,nofootinbib,showpacs,superscriptaddress]{revtex4-1}

\usepackage{amsmath,amsfonts,amssymb,bm}
\usepackage{dsfont}
\usepackage{graphicx}
\usepackage{color}
\usepackage{soul}
\usepackage{slashed}

\usepackage[mathscr,scaled=1.15]{urwchancal}
\DeclareFontFamily{OT1}{pzc}{}
\DeclareFontShape{OT1}{pzc}{m}{it}%
{<-> s * [1.15] pzcmi7t}{}
\DeclareMathAlphabet{\mathpzc}{OT1}{pzc}{m}{it}

\definecolor{purple}{rgb}{0.5,0,0.5}
\definecolor{blue}{rgb}{0.0,0,0.9}

\begin{document}

\title{Valence-quark distribution functions in the kaon and pion}

\author{Chen Chen}
\email{seracron@ustc.edu.cn}
\affiliation{Hefei National Laboratory for Physical Sciences at the Microscale,
University of Science and Technology of China, Hefei, Anhui 230026, P. R. China}
\affiliation{Institute for Theoretical Physics and Department of Modern Physics,\\
University of Science and Technology of China, Hefei, Anhui 230026, P. R. China}

\author{Lei Chang}
\email{lei.chiong@gmail.com}
\affiliation{School of Physics, Nankai University, Tianjin 300071, China}

\author{Craig D. Roberts}
\email{cdroberts@anl.gov}
\affiliation{Physics Division, Argonne National Laboratory, Argonne, Illinois 60439, USA}

\author{Shaolong Wan}
\email{slwan@ustc.edu.cn}
\affiliation{Institute for Theoretical Physics and Department of Modern Physics,\\ University of Science and Technology of China, Hefei, Anhui 230026, P. R. China}

\author{Hong-Shi Zong}
\email{zonghs@nju.edu.cn}
\affiliation{Department of Physics, Nanjing University, Nanjing 210093, China}

\date{29 January 2016}

\begin{abstract}
We describe expressions for pion and kaon dressed-quark distribution functions that incorporate contributions from gluons which bind quarks into these mesons and hence overcome a flaw of the commonly used handbag approximation.  The distributions therewith obtained are purely valence in character, ensuring that dressed-quarks carry all a meson's momentum at a characteristic hadronic scale and vanishing as $(1-x)^2$ when Bjorken-$x\to 1$.  Comparing such distributions within the pion and kaon, it is apparent that the size of $SU(3)$-flavour symmetry breaking in meson parton distribution functions is modulated by the flavour dependence of dynamical chiral symmetry breaking.
Corrections to these leading-order formulae may be divided into two classes, responsible for shifting dressed-quark momentum into glue and sea-quarks.  Working with available empirical information, we build an algebraic framework that is capable of expressing the principal impact of both classes of corrections.  This enables a realistic comparison with experiment which allows us to identify and highlight basic features of measurable pion and kaon valence-quark distributions.
We find that whereas roughly two-thirds of the pion's light-front momentum is carried by valence dressed-quarks at a characteristic hadronic scale, this fraction rises to 95\% in the kaon; evolving distributions with these features to a scale typical of available Drell-Yan data produces a kaon-to-pion ratio of $u$-quark distributions that is in agreement with the single existing data set; and predict a $u$-quark distribution within the pion that agrees with a modern reappraisal of $\pi N$ Drell-Yan data.
Precise new data are essential in order to validate this reappraisal and because a single modest-quality measurement of the kaon-to-pion ratio cannot be considered definitive.
\end{abstract}


\maketitle

\section{Introduction}
Light pseudoscalar mesons are of great interest in hadron and nuclear physics, in large part because they are the Nambu-Goldstone modes which arise as a consequence of dynamical chiral symmetry breaking (DCSB) and quite likely, therefore, play a material role in the realisation of confinement within the Standard Model.  It follows that comparisons between kaon and pion properties can reveal the relative impact of explicit and dynamical chiral symmetry breaking on hadron properties in an environment where those impacts are likely to be most heavily felt.  In proceeding toward an understanding of nonperturbative dynamics within quantum chromodynamics (QCD), it is important to expose and explain the effects of this interplay on the gluon and quark parton distribution functions (PDFs) within these hadrons.  Naturally, to be truly informative, feedback between experiment and theory is crucial \cite{Holt:2010vj}.

Experimental data on pion and kaon PDFs is, however, sparse.  It has only been obtained in mesonic Drell-Yan scattering from nucleons in heavy nuclei, with information on the pion's PDFs reported in Refs.\,\cite{Badier:1983mj, Betev:1985pg, Conway:1989fs} and results for the ratio of kaon and pion distribution functions presented in Ref.\,\cite{Badier:1980jq}.  Newer data is not available; but would be welcome, owing to persistent doubts about the large Bjorken-$x$ behaviour of the pion's valence-quark PDF \cite{Holt:2010vj} and because a single modest-quality measurement of the kaon-to-pion ratio cannot be considered definitive.  An approved experiment \cite{Keppel:2015, McKenney:2015xis}, using tagged deep inelastic scattering at the upgraded Jefferson Laboratory (JLab\,12), should contribute to a resolution of the pion question; and a similar technique might also serve for the kaon.  Such studies are part of an extensive and diverse range of hadron structure experiments planned at JLab\,12 \cite{Dudek:2012vr, Aznauryan:2012baS}.  Furthermore, new mesonic Drell-Yan measurements at modern facilities could yield valuable information on $\pi$ and $K$ PDFs \cite{Londergan:1995wp, Londergan:1996vh}, as could two-jet experiments at the large hadron collider \cite{Petrov:2011pg}; and, looking further ahead, an electron ion collider would be capable of providing access to pion and kaon structure functions through measurements of forward nucleon structure functions \cite{Holt:2000cv, Accardi:2012qut}.

The calculation of pion and kaon internal structure would at first sight seem to be straightforward, since these systems are bound-states involving just two valence quarks.  However, here first impressions are misleading owing to both the loss of particle number conservation in quantum field theory and DCSB, which together place severe constraints on any approach to the computation of pion and kaon properties.  A valid framework must be capable of simultaneously explaining these systems as conventional bound-states in quantum field theory and Nambu-Goldstone modes, with all the incumbent corollaries (\emph{e.g}.\ Refs.\,\cite{Maris:1997hd, Holl:2004fr, Holl:2005vu, Qin:2014vya}); and most models fail this test.  We will therefore address the problem using QCD's Dyson-Schwinger equations (DSE) \cite{Cloet:2013jya}, which are well suited to this exercise because of the existence of nonperturbative, symmetry-preserving truncation schemes \cite{Munczek:1994zz, Bender:1996bb, Chang:2009zb} that readily accommodate and explain the dichotomy of Nature's Nambu-Goldstone modes.  Early DSE predictions and results from other approaches are reviewed elsewhere \cite{Holt:2010vj}, and some more recent theoretical analyses are described in Refs.\,\cite{Frederico:2009fk, Nguyen:2011jy, Alberg:2011yr, Nam:2012vm, Bali:2013gya, Mezrag:2014jka, deMelo:2015yxk}.  Herein, we choose to follow Ref.\,\cite{Chang:2014lva} and develop an insightful perspective on kaon and pion PDFs using an algebraic framework that has also proven useful in studies of meson parton distribution amplitudes \cite{Shi:2015esa}.

Following this Introduction, Sec.\,\ref{secQDFs}, provides a little background to the measurement of PDFs and theoretical expectations.  In Sec.\,\ref{secComputing} we introduce formulae for meson valence dressed-quark PDFs, describe an algebraic framework that enables their straightforward computation and detail insights this provides.  Section~\ref{secBuilding} explains why, even at a typical hadronic scale, the valence dressed-quark structure of mesons as perceived in deep inelastic processes must be augmented by sea-quark and gluon contributions; and details a simple but realistic means of achieving this.  That positions us to describe comparisons with extant data in Sec.\,\ref{secDrawing} and reveal insights that such comparisons suggest.  We conclude in Sec.\,\ref{secEpilogue}.

\section{Quark distribution functions}
\label{secQDFs}
The hadronic tensor relevant to inclusive deep inelastic lepton-$M_5$ scattering, where $M_5$ is a pseudoscalar meson, may be expressed in terms of two invariant structure functions \cite{Jaffe:1985je}.  In the deep-inelastic Bjorken limit \cite{Bjorken:1968dy}: $q^2\to\infty$, $P\cdot q \to -\infty$ but $x:= - q^2/[2 P \cdot q]$ fixed,
that tensor can be written $(t_{\mu\nu} = \delta_{\mu\nu}-q_\mu q_\nu/q^2, P_\mu^{\,t}= t_{\mu\nu}P_\nu)$
\begin{subequations}
\begin{align}
W_{\mu\nu}(q;P) & = F_1(x)\, t_{\mu\nu} - \frac{F_2(x)}{P\cdot q}
\,P_\mu^{\,t} P_\nu^{\,t}\,,\\
F_2(x) & = 2 x F_1(x)\,.
\end{align}
\end{subequations}
$F_1(x)$ is the meson structure function, which provides access to its quark distribution functions:
\begin{equation}
\label{qPDF}
F_1(x) = \sum_{q\in M_5} \, e_q^2 \, [q^{M_5}(x)+\bar q^{M_5}(x)]\,,
\end{equation}
where $e_q$ is the quark's electric charge.  Bjorken-$x$ is equivalent to the light-front momentum fraction of the struck parton; and the structure function may be computed from the imaginary part of the virtual-photon--meson forward Compton scattering amplitude: $\gamma(q) + M_5(P) \to \gamma(q) + M_5(P)$ \cite{Landshoff:1970ff}.

The sum in Eq.\,\eqref{qPDF} runs over all quark flavours: in the $\pi^+$ it is dominated by $u^\pi(x)$, $\bar d^\pi(x)$, and in the $K^+$, by $u^K(x)$, $\bar s^K(x)$.  Notably, in the $\mathpzc{G}$-parity symmetric limit, which we employ throughout, $u^\pi(x)=\bar d^\pi(x)$.  On the other hand, one expects $u^K(x)\neq\bar s^K(x)$.  Indeed, the large difference between the current-masses of the $s$-quark and the $u$- and $d$-quarks should lead to some very interesting effects in the kaon structure function, \emph{e.g}.\ owing to its larger mass, the $s$-quark should carry more of the charged-kaon's momentum than the $u$ quark and hence the valence $u$-quark distribution in the kaon should peak at $x<1/2$.  Just how much less depends on the interplay between explicit and dynamical chiral symmetry breaking.

Here it is worth reiterating one of the earliest predictions of the QCD parton model, augmented by features of perturbative QCD (pQCD), \emph{i.e}.\ at large Bjorken-$x$ and at a scale characteristic of nonperturbative QCD, $\zeta_H$, the valence-quark distribution function in a pseudoscalar meson behaves as follows \cite{Ezawa:1974wm, Farrar:1975yb, Berger:1979du}:
\begin{equation}
\label{PDFQCD}
q^{M_5}(x;\zeta_H) \sim (1-x)^{2+\gamma},
\end{equation}
where $\gamma\gtrsim 0$ is an anomalous dimension.

Verification of Eq.\,\eqref{PDFQCD} will serve as an important milestone on the path toward confirmation of QCD as the theory of strong interactions.  In this connection we recall that Ref.\,\cite{Conway:1989fs} (the E615 experiment) reported a pion valence-quark PDF obtained via a leading-order (LO) pQCD analysis of their data, \emph{viz}.\ $u_V^\pi(x) \sim (1-x)$, seemingly a marked contradiction of Eq.\,\eqref{PDFQCD}.  Subsequent DSE computations \cite{Hecht:2000xa} confirmed Eq.\,\eqref{PDFQCD} and eventually prompted reconsideration of the E615 analysis, with the result that at next-to-leading order (NLO) and including soft-gluon resummation \cite{Wijesooriya:2005ir, Aicher:2010cb}, the E615 data can be viewed as being consistent with Eq.\,\eqref{PDFQCD}.  New data are essential in order to check this reappraisal of the E615 data and settle the controversy.

These remarks highlight the value of theoretical methods that possess a realistic connection with QCD and can provide precise, reliable results for parton distribution functions on the valence-quark domain.  Again, most models fail in this regard, \emph{e.g}.\ claiming agreement with the E615 data as a success and ignoring the conflict with Eq.\,\eqref{PDFQCD}.  Lattice-QCD, too, is challenged in this respect.  The standard methods provide access to only the lowest three moments of a given PDF, which are insufficient to test Eq.\,\eqref{PDFQCD} \cite{Holt:2010vj}; and contemporary implementations of a recently proposed alternative \cite{Ji:2013dva} do not overcome this drawback \cite{Lin:2014yra}.

\section{Computing valence-quark distribution functions}
\label{secComputing}
\subsection{Basic formulae}
In calculating valence dressed-quark distribution functions for the kaon and pion, we follow Ref.\,\cite{Chang:2014lva}, which demonstrated that the impulse-approximation (\emph{handbag}) expression typically used to define these distribution functions is deficient because it omits contributions from the gluons which bind quarks into the meson.  The corrected formula for the pion's valence $u$-quark distribution is:
\begin{align}
\nonumber
u_V^\pi&(x)  =  N_c  {\rm tr} \! \int_{dk}\!
\delta_n^x(k_\eta^\pi)\,
 \\
&\times \partial_{k_\eta^\pi} \left[ \Gamma_\pi(k_\eta^\pi,-P_\pi)  S(k_\eta^\pi)\right] \Gamma_\pi(k_{\bar\eta}^\pi,P_\pi)\, S(k_{\bar\eta}^\pi)\,, \label{uvpix}
\end{align}
where the derivative acts only on the bracketed terms.  Here, $N_c=3$; the trace is over spinor indices; $\int_{dk}$ represents a translationally invariant regularisation of the four-dimensional momentum integral; $\delta_n^{x}(k_\eta^\pi):= \delta(n\cdot k_\eta^\pi - x n\cdot P_\pi)$; $n$ is a light-like four-vector, $n^2=0$; $P_\pi$ is the pion's four-momentum, $P_\pi^2=-m_\pi^2$ and $n\cdot P_\pi = -m_\pi$, with $m_\pi$ being the pion's mass; and $k_\eta^\pi = k + \eta P_\pi$, $k_{\bar\eta}^\pi = k - (1-\eta) P_\pi$, $\eta\in [0,1]$.
The two functions appearing in Eq.\,\eqref{uvpix} are the dressed-quark propagator:
\begin{equation}
\label{DressedSk}
S_f(k) = Z_f(k^2)/[i \gamma\cdot k + M_f(k^2)]\,,
\end{equation}
where we have added a flavour label, $f=u,s$; and the meson Bethe-Salpeter amplitude:
\begin{align}
\nonumber \Gamma_M(k;P) & = \gamma_5 \big[ i E_M(k;P) + \gamma\cdot P F_M(k;P)  \\
& + \gamma\cdot k G_M(k;P) + \sigma_{\mu\nu} k_\mu P_\nu H_M(k;P) \big]\,.
\end{align}
 Owing to Poincar\'e covariance, no observable can legitimately depend on the definition of the relative momentum, \emph{i.e}.\ $\eta$;
and recall $u^\pi = \bar d^\pi$ in the $\mathpzc{G}$-parity limit considered herein.

We note that Eq.\,\eqref{uvpix} was derived using the rainbow-ladder truncation, which is the leading term in any systematic DSE truncation scheme \cite{Binosi:2016rxz}.  However, as we shall see, it can also serve as a symmetry-preserving definition of the valence-quark PDF that can then be employed with any reasonable forms for the dressed-quark propagators and Bethe-Salpeter amplitudes.

Analogous formulae for the kaon PDFs are:
\begin{subequations}
\label{valenceqK}
\begin{align}
\nonumber u_V^K&(x) =  N_c  {\rm tr} \! \int_{dk}\!
\delta_n^x(k_\eta)\,\\
&\times \partial_{k_\eta} \left[ \Gamma_K(k_\eta,-P)  S_u(k_\eta)\right] \Gamma_K(k_{\bar\eta},P)\, S_s(k_{\bar\eta})\,, \label{uvKx}\\
\nonumber \bar s_V^K&(x)  =  N_c  {\rm tr} \! \int_{dk}\!
\delta_n^x(k_\eta)\,\\
&\times \Gamma_K(k_{\eta},-P)\, S_u(k_{\eta})\partial_{k_{\bar\eta}} \left[ \Gamma_K(k_{\bar\eta},P)  S_s(k_{\bar\eta})\right] \,, \label{svKx}
\end{align}
\end{subequations}
where here $P$ refers to the kaon total momentum, so that $P^2=-m_K^2$, $n\cdot P = -m_K$.

\subsection{Algebraic framework}
In order to complement insights drawn about the dressed-quark structure of the pion in Ref.\,\cite{Chang:2014lva}, we compute the valence dressed-quark PDFs using the following elements ($\Delta_{M_f}(s)=1/[s+M_f^2]$) \cite{Chang:2013pq}:
\begin{subequations}
\label{NakanishiASY}
\begin{align}
\label{eq:sim1}
S_f(k) & = [-i\gamma\cdot k+M_f]\Delta_{M_f}(k^2)\,,\\
%
%
%
\nonumber
\Gamma_\pi(k_{\bar\eta/\eta};\pm P)&= i \gamma_5\frac{M_u}{{\mathpzc{n}}_\pi } \frac{3}{4}\int dz\\
& \times  (1-z^2) M_u^2 \Delta_{\Lambda_\pi}(k^2_{z}) \,, \\
%
\nonumber \Gamma_K(k_{\bar\eta/\eta};\pm P)&= i\gamma_5\frac{M_R}{{\mathpzc{n}}_K} \frac{3}{4}\int dz  \\
& \times (1-z^2)(1+\beta z) M_{us}^2 \Delta_{\Lambda_K}(k^2_{z})\,, \label{GammaK}
\end{align}
\end{subequations}
where $M_{us}^2 = M_u M_s$, $M_R=M_{us}^2/[M_u+M_s]$, $k_{z}=k_{\bar\eta/\eta} + (z \pm 1) P/2$, and $\beta$ is a skewing parameter that grows with the $s$-$u$ mass difference and serves to deform the kaon Bethe-Salpeter amplitude so that it exhibits a realistic asymmetry between the $u$- and $\bar s$-quarks.  Naturally, the normalisation constants $\mathpzc{n}_{\pi,K}$ are not parameters.  They are defined via the canonical procedure, \emph{i.e}.\ by requiring that meson bound-states contain one valence-quark and one valence-antiquark:
\begin{subequations}
\begin{align}
1 & = \int_0^1dx \, u_V^\pi(x)   \,,\\
1 & = \int_0^1 dx u_V^K(x) = \int_0^1 dx \bar s_V^K(x) \,.
\end{align}
\end{subequations}
It is worth remarking here that calculations using solutions of realistic gap and Bethe-Salpeter equations are planned; but whilst they will complement and extend Ref.\,\cite{Nguyen:2011jy}, they cannot materially alter the conclusions that our analysis using Eqs.\,\eqref{NakanishiASY} will subsequently enable to be drawn.  Most importantly, perhaps, such studies might enable improved constraints to be placed on sea-quark and gluon distributions within mesons, which are poorly known at present.

Using the algebraic formulae in Eqs.\,\eqref{NakanishiASY}, working in the limit of $u$- and $d$-quarks with zero current-mass, so that $P_\pi^2=-m_\pi^2=0$, and with $\Lambda_\pi=M_u$, one obtains the following algebraic expression for the pion's chiral-limit valence-quark distribution \cite{Chang:2014lva}:
\begin{align}
\nonumber
u_{V}^0&(x) =
    \frac{72}{25} \big[ x^3 (x [2 x-5]+15) \ln (x) + (x [2 x+1]+12) \\
& \times (1-x)^3 \ln (1-x) +2 x (6-[1-x] x) (1-x)\big].\!\!
\label{qRLSymmetric}
\end{align}
This function is symmetric about $x=1/2$ and consequently, without tuning,
\begin{equation}
\label{symmuvpi}
\langle x \rangle_u^0 = \int_0^1 dx\, x \, u_{V}^0(x) = \frac{1}{2}\,.
\end{equation}
That is logical because the dressed-quark and -antiquark are the sole measurable constituents of the pion in any internally-consistent computation using a standard dressed-quark basis: at the hadronic scale, they and their associated bound-state amplitude absorb and contain all contributions from sea or glue partons.  It follows that if the dressed-quark carries a fraction $x$ of the pion's momentum, the dressed-antiquark carries $[1-x]$.  In addition, one readily finds
\begin{equation}
\label{valencepower}
u_{V}^0(x) \stackrel{x\simeq 1}{=} \frac{216}{5}\, (1-x)^2 + \mbox{O}([1-x]^3)\,,
\end{equation}
which is the power-law predicted by the QCD parton model, obtained simply and exactly.  Owing to symmetry under $x\leftrightarrow [1-x]$, the same power-law is manifest on $x\sim 0$, a result which emphasises that $u_{V}^0(x)$ is truly a constituent-like distribution: any sea-quark or gluon contamination would produce a marked asymmetry.
Notably, the symmetry is a property of Eq.\,\eqref{uvpix}: it is found irrespective of the forms used for the quark propagator and Bethe-Salpeter amplitudes; and it is this feature which justifies our use of Eqs.\,\eqref{uvpix}, \eqref{valenceqK} as practical definitions of the valence dressed-quark PDFs.

The quantities $M_{u,s}$, $\Lambda_{\pi,K}$ in Eqs.\,\eqref{NakanishiASY} are, respectively, dressed-quark mass and meson Bethe-Salpeter amplitude width parameters.  We choose $M_u=0.4\,$GeV because this value provides a good description of the pion's elastic electromagnetic form factor in the generalised parton distribution analysis of Ref.\,\cite{Mezrag:2014jka}; set $M_s= 1.2 M_u$, which is typical of the value obtained for the ratio of Euclidean constituent-quark masses in phenomenologically efficacious DSE analyses \cite{Maris:1997tm, Shi:2015esa}; and float $\Lambda_{\pi,K}$ to fit the leptonic decay constants of these pseudoscalar mesons:
\begin{subequations}
\label{fHformulae}
\begin{align}
&f_\pi = \frac{N_c}{n\cdot P_\pi} {\rm tr_D} \! \int_{dk}\!\!\!
\gamma_5 \gamma\cdot n S_u(k_\eta^\pi)\Gamma_\pi (k_\eta^\pi;P_\pi)S_d(k_{\bar \eta}^\pi)\,,\\
& f_K = \frac{N_c}{n\cdot P} {\rm tr_D}\int_{dk}
\gamma_5 \gamma\cdot n S_u(k_\eta)\Gamma_K (k_\eta;P)S_s(k_{\bar \eta})\,,
\end{align}
\end{subequations}
where $m_\pi=0.14\,$GeV, $m_K=0.49\,$GeV.  With $\Lambda_\pi = 0.52 M_u$, $\Lambda_K=0.93 M_u$, we obtain $f_\pi =0.092\,$GeV, $f_K = 0.11\,$GeV, in agreement with experiment \cite{Agashe:2014kda}.

\subsection{Valence dressed-quark distributions}
Using the parameter values just described cannot alter the power-law behaviour of the computed PDFs in the neighbourhood of the endpoints: $x=0,1$.  Hence, as with $u_V^0(x)$, our complete results for $u_{V}^\pi(x)$, $u_{V}^K(x)$, $\bar s_{V}^K(x)$ must also conform with the QCD prediction, Eq.\,\eqref{PDFQCD}.

On the other hand, with the additional complexity in Eqs.\,\eqref{NakanishiASY} and nonzero values for the meson masses, it is difficult to obtain algebraic forms for the dressed-quark PDFs.  We therefore adopt a different approach.
Namely, given the known endpoint behaviour of the valence-quark PDFs at the hadronic scale, it is plain that they have the following representation:
\begin{equation}
\label{Gegenbauer}
q_V(x) = 30 [x(1-x)]^2 \bigg[ 1+ \sum_{j=1}^{j_m} a_j^{5/2} C_j^{5/2}(2x-1) \bigg]\,,
\end{equation}
where $\{C_j^{5/2}|j=1,2,\ldots \}$ are those Gegenbauer polynomials which form a complete, orthonormal set with respect to the measure $ [x(1-x)]^2$.  We therefore compute the moments of the distributions:
\begin{align}
& \langle x^m \rangle_u^\pi  = \int_0^1 dx \, x^m \, u_V^\pi(x) \\
\nonumber
& = \frac{N_c}{n\cdot P_\pi} {\rm tr_D} \int_{dk} \, \left[\frac{n\cdot k_\eta^\pi}{n\cdot P_\pi}\right]^m
  \\
& \quad \quad \times \partial_{k_\eta^\pi} \left[ \Gamma_\pi(k_\eta^\pi,-P_\pi)  S(k_\eta^\pi)\right]  \Gamma_\pi(k_{\bar\eta}^\pi,P_\pi)\, S(k_{\bar\eta}^\pi)\,,
\end{align}
with analogous expressions for $\langle x^m \rangle_u^K$, $\langle x^m \rangle_{\bar s}^K$; and vary the coefficients $\{a_j^{5/2}|\,j=1,2,\ldots \}$ in Eq.\,\eqref{Gegenbauer} so as to obtain a best least-squares fit to those calculated  moments.

This reconstruction procedure converges very quickly, \emph{e.g}.\ using just the first nontrivial moment of Eq.\,\eqref{qRLSymmetric} to determine $a_2^{5/2}$, with all other coefficients set to zero, one obtains a curve via Eq.\,\eqref{Gegenbauer} that is visually indistinguishable from the exact result.  Applied to the distributions $u_{V}^\pi(x)$, $u_{V}^K(x)$, $\bar s_{V}^K(x)$ obtained using Eqs.\,\eqref{NakanishiASY} and the parameter values determined via Eqs.\,\eqref{fHformulae}, it yields:
\begin{equation}
\label{PDFGegenbauer}
\begin{array}{lccccc}
                & a_1 & a_2 & a_3 & a_4 & a_5 \\
u_V^\pi & 0 & -0.0382 & 0 & 0 & 0 \\
u_V^K & -0.175 & -0.0181 & \phantom{-}0.0101 & 0.0012 & -0.0011\\
\bar{s}_V^K &  \phantom{-}0.175 & -0.0181 & -0.0101 & 0.0012 & \phantom{-}0.0011
\end{array}\,.
\end{equation}

\begin{figure}[t]
\centerline{\includegraphics[width=0.45\textwidth]{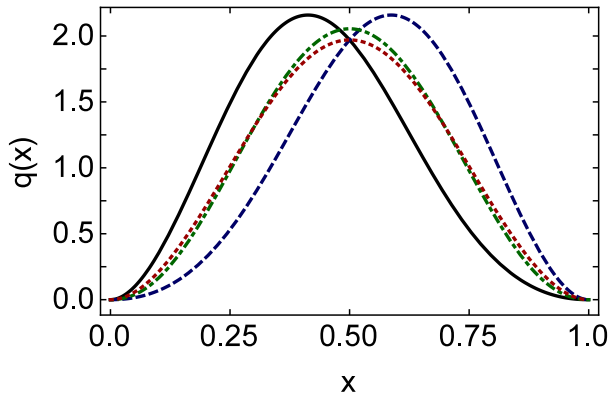}}
\caption{\label{figPDFsQ0}  Valence-quark PDFs at the hadronic scale, $\zeta_H$, defined by Eqs.\,\eqref{Gegenbauer}, \eqref{PDFGegenbauer}:
$u_V^K(x)$, solid (black) curve;
${\bar s}_V^K(x)$, dashed (blue) curve;
$u_V^\pi(x)$, dot-dashed (green) curve; and
$[u_V^K(x)+{\bar s}_V^K(x)]/2$ dotted (red) curve.
}
\end{figure}

The distributions defined by Eq.\,\eqref{PDFGegenbauer} are depicted in Fig.\,\ref{figPDFsQ0}.  We discuss their evolution with resolving scale, $\zeta$, below; but here it is worth noting that the following ratios are $\zeta$-independent and hence are a discriminating probe of the nonperturbative dynamics \cite{Holt:2010vj}:
\begin{equation}
\label{xoneratios}
\left. \frac{u_V^K(x)}{u_V^\pi(x)}\right|_{x\to 1} = 0.37\,, \quad
\left. \frac{u_V^\pi(x)}{\bar{s}_V^K(x)}\right|_{x\to 1} = 0.29\,.
\end{equation}

The curve $[u_V^K(x)+\bar s_V^K(x)]/2$ is also drawn in the figure:  it is symmetric under $x\leftrightarrow (1-x)$.  That feature outcome is an obvious consequence of the definitions, Eqs.\,\eqref{valenceqK}, and a requirement of any expressions for the valence dressed-quark PDFs if one is to guarantee momentum conservation without tuning, \emph{i.e}.\
\begin{equation}
\label{kaonmom}
\int_0^1 dx \, x [ u_V^K(x) + \bar s_V^K(x) ] = 1\,,
\end{equation}
independent of model details and any associated parameter values.  This curve is similar to $u_V^\pi(x)$ but not identical, \emph{e.g}.\  $[ u_V^K(x) + \bar s_V^K(x) ] /u_V^\pi(x) = 1.89$ at $x=1$.

Another important feature of the computed valence-quark distributions is the shift in the peak of $u_V^K(x)$ away from $x=0.5$ or, equivalently, the analogous shift in $\bar s_V^K(x)$.  These quantities measure the scale of $SU(3)$-flavour symmetry breaking and indicate its origin; and with the distributions depicted in Fig.\,\ref{figPDFsQ0} one finds a shift of 17\%.  This result is nearly identical to the 16\% shift in the peak of the leading-twist $s$-quark parton distribution amplitude (PDA) in the kaon \cite{Shi:2015esa}.  Additionally, it almost matches the relative shift in dressed-quark masses, \emph{i.e}.\ $M_s$ with respect to $M_u$.  Consequently, in valence dressed-quark PDFs it is the flavour-dependence of DCSB that modulates the strength of $SU(3)$-flavour symmetry breaking, which is therefore far smaller than one would na\"{\i}vely expect based on the difference between the current-masses of $s$- and $u$-quarks.  This is true of numerous other quantities, \emph{e.g}.\ the ratio of neutral- and charged-kaon electromagnetic form factors measured in $e^+e^-$ annihilation at $s_U=17.4\,$GeV$^2$, $|F_{K_S K_L}(s_U)|/|F_{K_- K_+}(s_U)|\approx 0.12$ \cite{Seth:2013eaa}; and the leptonic decay constant ratios $f_K/f_\pi$, $f_{B_s}/f_B=1.21$ \cite{Ivanov:2007cw, Christ:2014uea}.  

\section{Building realistic distributions}
\label{secBuilding}
\subsection{Including sea-quarks and glue}
The dressed-quark basis employed hitherto provides a good description of a wide range of hadron properties \cite{Cloet:2013jya}; and it yields purely valence-quark distributions.  This last quality is evident in the derivation of Eq.\,\eqref{uvpix} presented in Ref.\,\cite{Chang:2014lva}, which also explains that their are corrections to Eq.\,\eqref{uvpix} (and Eqs.\,\eqref{valenceqK}, its analogues for the kaon), which can be separated into two classes: [C1] redistributes baryon-number and momentum into the dressed-quark sea; and [C2] shifts momentum into the dressed-gluon distribution within the meson.

Some obvious contributions within [C1] are those associated with what may be called \emph{resonant} or \emph{meson-cloud} corrections to the kernels in the gap and scattering equations, simple examples of which are
\begin{subequations}
\label{egC1}
\begin{align}
\nonumber
\pi&^+ = u \bar d \\
& \to u (\bar d d) \bar d = (u \bar d) (d \bar d) \sim \pi^+ \rho^0 \to u \bar d = \pi^+,\\
\nonumber
K&^+  = u \bar s \\
& \to u (\bar u u) \bar s = (u \bar u) (u \bar s) \sim \pi^0 K^{\ast +} \to u \bar s = K^+.
\label{Kmesonloop}
\end{align}
\end{subequations}
These sequences describe \emph{bare}-\emph{mesons}, built in a dressed-quark basis, adding additional structural components to their Bethe-Salpeter wave functions to produce the fully-dressed and hence physical state.  Such processes enable the hard photon to interact with sea-quark components of the physical meson, thereby shifting momentum into a sea-quark distribution within the meson.

Following Ref.\,\cite{Chang:2014lva}, let us first consider the pion and associate a total flux ``$Z_\pi$'' with such fluctuations.  In a symmetry preserving treatment, such processes do not change the total baryon-number content of the pion but they do reduce the probability of finding the bare-pion within the physical pion; and hence the quark distribution becomes
\begin{equation}
u_{Vs}^\pi(x) = (1-Z_\pi) u_V^\pi(x) + Z_\pi u_M^\pi(x) \,,
\end{equation}
where $u_{M}^\pi(x)$ describes the cumulative effect on the PDF of all resonant corrections to the bound-state wave function computed in the dressed-quark basis and \mbox{$\int_0^1dx\,u_{M}^\pi(x)=1$}.

In order to determine $Z_\pi$, we note that with realistic masses, meson-loop corrections to the pion electromagnetic form factor at $Q^2=0$ are an O(5\%) effect.  This is evident in Ref.\,\cite{Alkofer:1993gu} and also in the result that, in the chiral limit, the pion's leptonic decay constant is \cite{Gasser:1983yg} $f_0^2 \approx (0.09\,{\rm GeV})^2$ cf.\ experiment \cite{Agashe:2014kda} $f_\pi^2 \approx (0.092\,{\rm GeV})^2$.  As in Ref.\,\cite{Chang:2014lva}, we therefore fix
\begin{equation}
\label{Zpivalue}
Z_\pi(\zeta_H)=0.05\,.
\end{equation}

One must now decide upon the value of $Z_K$.  Eq.\,\eqref{Kmesonloop} indicates the lightest possible intermediate state.  An alternative, equally simple contribution involves $u(\bar s s)\bar s \sim K^+ \phi^0$, which is a more massive combination, whereas the analogous term for the pion is $\sim \pi^0 \rho^+$.   These observations indicate that the pion-to-kaon ratio of mass-squared denominators is roughly one-third and hence $Z_K \sim Z_\pi/3$.  We therefore set
\begin{equation}
\label{ZK0}
Z_K(\zeta_H) = 0
\end{equation}
because our subsequent analysis cannot reasonably be expected to exhibit a reliable sensitivity to 1\% effects.

The dressed-quark structure of mesons receives corrections in addition to those exemplified in Eqs.\,\eqref{egC1}.  Namely, one may readily identify corrections to Eqs.\,\eqref{uvpix}, \eqref{valenceqK} that shift momentum into the meson's gluon distribution \cite{Chang:2014lva}, \emph{e.g}.\ one can draw diagrams in which the struck dressed-quark carries a fraction $x$ of the meson's momentum, but the momentum of the spectator system is shared between the dressed-antiquarks and -gluons: attributing a net $x_g>0$ to the dressed-gluon, then the dressed-antiquark carries $1-x-x_g$.  In a symmetry preserving treatment, these and other corrections in [C2] have no impact on net baryon number within the meson but they do rob momentum from the baryon-number-carrying dressed-partons; namely, $q_{V,M}^{\pi,K}(x) \to q_{V_g,M_g}^{\pi,K}(x)$, with
\begin{subequations}
\label{DefineShift}
\begin{eqnarray}
\mbox{$\int_0^1$} dx \,q_{V_g,M_g}^{\pi,K}(x) \;\; &=& \mbox{$\int_0^1$} dx \,q_{V,M}^{\pi,K}(x)\,,\\
\mbox{$\int_0^1$} dx \, x\,q_{V_g,M_g}^{\pi,K}(x) &<& \mbox{$\int_0^1$} dx \, x\, q_{V,M}^{\pi,K}(x)\,,
\end{eqnarray}
\end{subequations}
where $q$ represents $u$ and/or $\bar s$ as appropriate.  Thus, with $\delta_g q_{V,M}^{\pi,K}(x) := q_{V_g,M_g}^{\pi,K}(x)-q_{V,M}^{\pi,K}(x)$, one arrives finally at expressions for the complete dressed-quark distribution functions at the hadronic scale:
\begin{subequations}
\label{qPDFfinal}
\begin{align}
\nonumber
u^\pi(x) & = (1-Z_\pi)[u_{V}^\pi(x) + \delta_g u_{V}^\pi(x)]\\
& \quad + Z_\pi[u_{M}^\pi(x) + \delta_g u_{M}^\pi(x)] \,, \label{uPDFfinal}\\
\nonumber
u^K(x) & = (1-Z_K)[u_{V}^K(x) + \delta_g u_{V}^K(x)]\\
& \quad + Z_K[u_{M}^K(x) + \delta_g u_{M}^K(x)] \,,\\
\nonumber
\bar s^K(x) & = (1-Z_K)[\bar s_{V}^K(x) + \delta_g \bar s_{V}^K(x)]\\
& \quad + Z_K[\bar s_{M}^K(x) + \delta_g \bar s_{M}^K(x)] \,.
\end{align}
\end{subequations}

A procedure one may follow in order to compute a meson's valence-quark distribution function, Eq.\,\eqref{qPDFfinal}, is now apparent: begin with results obtained in the dressed-quark basis using sophisticated kernels for those equations involved in bound-state calculations and with the resolution set via renormalisation at a particular scale, $\zeta_H$; then proceed systematically to add the corrections identified above; and, finally, use DGLAP evolution \cite{Dokshitzer:1977, Gribov:1972, Lipatov:1974qm, Altarelli:1977} to obtain the result at any other scale $\zeta>\zeta_H$.  The last step is simply a labour-saving device because it eliminates the need for complete recomputation of the PDF at the new scale.  In this way one fixes \emph{a priori} that parameter, $\zeta_H$, which practitioners usually identify as the \emph{typical hadronic scale}, 
and whose variation provides them with considerable flexibility as they seek to validate their model through a fit to data.

In this connection, one might ask for the value of $\zeta_H$ at which the result computed using the dressed-quark basis alone should be most realistic.  That is $\zeta_H\simeq 0\,$GeV, because the light-front momentum fraction carried by dressed-sea and -glue diminishes as $\zeta$ is reduced.  However, use of the available DGLAP equations at such a small value of $\zeta_H$ is impossible because they are only valid on the perturbative domain.  What, then, is a suitable compromise?  An answer was provided in Ref.\,\cite{Holt:2010vj}: one should use $\zeta_H \geq 2\Lambda_{\rm QCD} \approx 0.5\,$GeV, which corresponds to a scale whereat the chiral-limit dressed-quark mass-function, $M(k^2)$ in Eq.\,\eqref{DressedSk}, is concave-up (convex) and dropping rapidly but does not yet exhibit the behaviour associated with its truly asymptotic momentum-dependence.  As explained elsewhere \cite{Chang:2012rk}, it is only for momenta within this domain that a rigorous connection with pQCD exists: it is impossible to begin at a smaller scale because then the crucial elements in any calculation, \emph{e.g}.\ the dressed-quark propagator, exhibit momentum dependence that is essentially nonperturbative in origin.  Notably, the expansion parameter in the DGLAP equations is $\alpha(s)/[2\pi]$, where $\alpha(s)$ is the strong running coupling; and  $\alpha(4\Lambda_{\rm QCD}^2)/(2\pi)\approx 0.17$ whereas $\alpha(2\Lambda_{\rm QCD}^2)/(2\pi)\approx 0.34$, which further vitiates any choice $\zeta_H<2\Lambda_{\rm QCD}$.

Some additional remarks are in order here.  Notwithstanding the existence of calculable corrections to results obtained using the dressed-quark basis, that basis provides a good foundation for describing numerous hadron observables.  This is readily illustrated via the pion's electromagnetic form factor, $F_\pi(Q^2)$.  Meson-loop corrections only measurably affect its low-$Q^2$ behaviour, contributing $\lesssim 15\%$ to $r_\pi^2$ (squared-charge-radius) \cite{Alkofer:1993gu}; and gluonic corrections serve only to modify the form-factor's anomalous dimension \cite{Lepage:1980fj, Maris:1998hc, Cloet:2013jya}.  The salient features of $F_\pi(Q^2)$, including parton model scaling and the existence of scaling violations, are captured in the dressed-quark basis \cite{Chang:2013nia}.

\subsection{Explicating sea-quark and glue distributions}
Owing to Eq.\,\eqref{ZK0} and the associated discussion, it is only necessary to specify the profile of the pion's sea-quark distribution; and in this we draw guidance from empirical information on $\pi N$ Drell-Yan \cite{Gluck:1999xe}:
\begin{equation}
\label{GRSsea}
x u_M^\pi(x) = \frac{1}{\mathpzc{n}} x^{\bar\alpha} (1-x)^{\bar\beta} (1 - \bar\gamma \sqrt{x} + \bar\delta  x)
\end{equation}
where $1/\mathpzc{n}$ is a simple algebraic factor that ensures $\int_0^1dx\,u_M^\pi(x)=1$.  Then, at $\zeta_H=0.51\,$GeV an empirical assessment of the pion's sea-quark distribution is consistent with
\begin{equation}
\label{SeaParameters}
\bar\alpha=0.16\,,\; \bar\beta=5.20\,,\;\bar\gamma=3.243\,,\;\bar\delta=5.206\,.
\end{equation}

The same consideration of $\pi N$ Drell-Yan shows that 29\% of the pion's momentum is carried by glue at $\zeta_H$ $[\langle x_g\rangle =0.29]$, in a distribution that has \cite{Gluck:1999xe} $\alpha_g \approx 3/2$ and $\beta_g \approx 1+\beta_V$, where $\beta_V$ is the exponent which characterises the pion's valence-quark distribution on $x\simeq 1$.  In Eq.\,\eqref{uPDFfinal}, we therefore emulate Ref.\,\cite{Chang:2014lva} and use $\delta_g u_{V,M}^\pi = \delta_g u^\pi$,
\begin{equation}
\label{shift}
\delta_g u^\pi(x) = \mathpzc{c}_g^\pi \, x^{\alpha_g-1} (1-x)^{\beta_g} P_1^{(\beta_g \, \alpha_g-1)}(2x-1)\,,
\end{equation}
with $\mathpzc{c}_g^\pi$ a parameter and $P_1$ a Jacobi polynomial, in order to shift 29\% of the dressed quarks' momentum into the gluon distribution.  [Equation~\eqref{shift} is consistent with Eqs.\,\eqref{DefineShift}.]  With $\beta_g=3$, owing to Eq.\,\eqref{valencepower}, one finds
\begin{equation}
\mathpzc{c}_g^\pi(\zeta_H)=8.50\,.
\end{equation}

All parameters in $u^\pi(x)$ are now fixed, so that the result we subsequently describe is a prediction for this distribution.  We are not so fortunate with the kaon: there are no published constraints on its gluon distribution.  We therefore employ Eq.\,\eqref{shift} for the kaon's gluon profiles, use $\mathpzc{c}_g^{K_u}(\zeta_H)$ as a parameter to be determined by fitting extant Drell-Yan data on the ratio $u^K(x)/u^\pi(x)$, and thereby provide a constraint on the fraction of the kaon's momentum carried by glue at the hadronic scale.  In order to proceed we must fix  $\mathpzc{c}_g^{K_s}(\zeta_H)$, which we do by requiring that gluons remove the same fraction of momentum from $u$- and $\bar s$-quarks in the kaon, \emph{viz}.
\begin{equation}
\frac{u^K(x)}{\bar s^K(x)} = \frac{u_V^K(x)}{\bar s_V^K(x)} \quad \Rightarrow \quad \mathpzc{c}_g^{K_s} = 1.29\,\mathpzc{c}_g^{K_u}.
\end{equation}
At this point, we have just one free parameter in our predictions for $u^\pi(x)$, $u^K(x)$, $\bar s^K(x)$, \emph{i.e}.\ $\mathpzc{c}_g^{K_u}$.

\section{Drawing comparisons with data}
\label{secDrawing}
\begin{figure}[t]
\centerline{\includegraphics[width=0.45\textwidth]{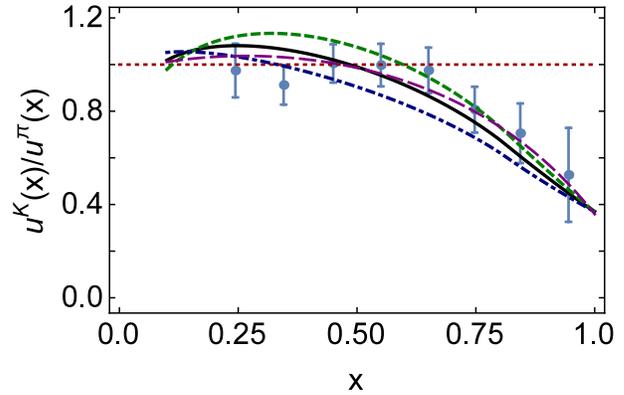}}
\caption{\label{figuKusratio}  $u^K(x)/u^\pi(x)$ at $\zeta=5.2\,$GeV:
solid (black) curve, obtained via LO evolution from $\zeta_H=0.51\,$GeV assuming 5\% of the kaon's momentum is carried by glue at this hadronic scale solid (black) curve; dashed (green) curve,  zero momentum carried by gluons; and dot-dashed (blue) curve, 10\% of the kaon's momentum carried by glue.
For comparison, an analysis of $\pi N$ Drell-Yan data suggests that 29\% of the pion's momentum is carried by glue at $\zeta_H$, as explained in connection with Eq.\,\eqref{shift}.
The long-dashed (purple) curve is the DSE prediction in Ref.\,\cite{Nguyen:2011jy}, obtained using numerical solutions of realistic gap and Bethe-Salpeter equations.
(Data in this figure are from Ref.\,\cite{Badier:1980jq}.  The dotted (red) line marks a value of unity for the ratio.  It is drawn to highlight the domain upon which one might be confident empirically that $u^K(x)/u^\pi(x)\neq 1$, \emph{viz}. $x\gtrsim 0.8$.)
}
\end{figure}

All that is required to report results for the valence-quark distribution in the pion is now specified.  However, in order to supply results for the kaon PDFs, the parameter $\mathpzc{c}_g^{K_u}$ must be determined.  In order to achieve that, we use leading-order DGLAP evolution from $\zeta_H=0.51\,$GeV to $\zeta_{5.2}=5.2\,$GeV and require a least-squares fit to the kaon-to-pion ratio of Drell-Yan cross-sections obtained from a sample of dimuon events with invariant mass $4.1 < M < 8.5\,$GeV \cite{Badier:1980jq}.  (\emph{N.B}.\ We choose $\zeta_{5.2}$ because that is the average mass for data taken in the E615 experiment \cite{Conway:1989fs, Wijesooriya:2005ir}, which covered bins with $4.05 < M < 8.53\,$GeV.)  In this way, one finds
\begin{equation}
\mathpzc{c}_g^{K_u}(\zeta_H) = 1.28 \quad \Rightarrow \quad \langle x_g \rangle^{K_u}(\zeta_H)=0.05\,,
\end{equation}
and the result depicted in Fig.\,\ref{figuKusratio}.  The evolved distributions may satisfactorily be interpolated by the following expression:
\begin{equation}
\label{xqxfit}
x q(x) = A x^\alpha (1-x)^\beta (1 - \gamma \sqrt{x} + \delta x)\,,
\end{equation}
with
\begin{equation}
\label{xqxfitparams}
\begin{array}{l|ccccc}
\zeta_{5.2}  & A &   \alpha  & \beta & \gamma  & \delta \\\hline
xu^\pi   &   \phantom{2}1.08   & 0.70   & 2.93   & 0\phantom{.86} & 5.48 \\
xu^K     &   18.62 & 1.56   & 2.93   & 0.86 & 0\phantom{.25}  \\
x\bar s^K     &   20.17 & 1.64   & 2.93   & 2.09 & 2.25
\end{array}\,.
\end{equation}
There is a marked similarity between our result (solid, black curve), obtained using simple algebraic inputs, and the DSE prediction in Ref.\,\cite{Nguyen:2011jy} (long-dashed, purple curve), which was computed using numerical solutions of realistic gap and Bethe-Salpeter equations.  This confluence suggests that the theoretical prediction of the ratio and explanation of its behaviour are sound, and argues strongly for empirical verification of the first and only experimental result \cite{Badier:1980jq}.
In connection with these predictions, it is important to remark that any differences generated by next-to-leading-order (NLO) evolution are readily masked by a 25\% increase in $\zeta_{\rm H}$ \cite{Gluck:1999xe} and are thus immaterial.


It is apparent in Fig.\,\ref{figuKusratio} that $\lim_{x\to 1} u^K(x)/u^\pi(x)$ is independent of the kaon's gluon (and sea) content at $\zeta_H$.  This feature of the ratio at $x=1$ is a corollary of its $\zeta$-independence, explained in connection with Eqs.\,\eqref{xoneratios}.  On the other hand \cite{Chang:2010xs}:
\begin{equation}
\label{xtoone}
\lim_{x\to 0} \frac{u^K(x;\zeta)}{u^\pi(x;\zeta)} \stackrel{\Lambda_{\rm QCD}/\zeta \simeq 0}{\to} 1 \,.
\end{equation}
This owes to inexorable growth in both mesons' gluon and sea-quark content driven by pQCD splitting mechanisms.  That content finally comes to overwhelm nonperturbatively generated differences between the internal structure of the pion and kaon.  The result in Eq.\,\eqref{xtoone} is analogous to the convergence of all meson PDAs to the conformal form as $\Lambda_{\rm QCD}/\zeta \to 0$ \cite{Lepage:1979zb,Efremov:1979qk,Lepage:1980fj}.

\begin{figure}[t]
\centerline{\includegraphics[width=0.45\textwidth]{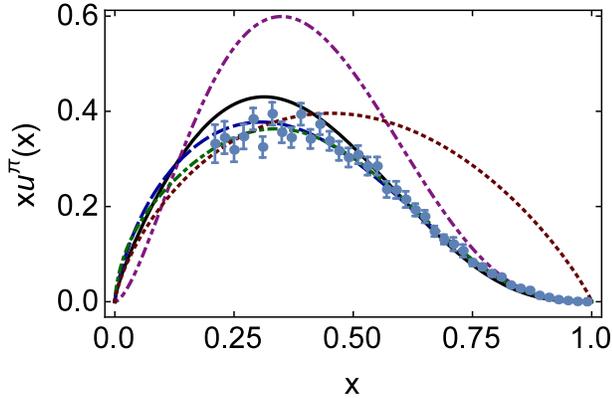}}
\caption{\label{figxuvpi}  $x u^\pi(x;\zeta_{5.2})$.  Solid (black) curve, our prediction, expressed in Eqs.\,\eqref{xqxfit}, \eqref{xqxfitparams};
dot-dot-dashed (purple) curve, result obtained when sea-quark and gluon contributions are neglected at $\zeta_H$, \emph{i.e}.\ using $u_V^\pi(x)$ from Eqs.\,\eqref{Gegenbauer}, \eqref{PDFGegenbauer};
dashed (blue) curve first DSE prediction \cite{Hecht:2000xa};
and data, Ref.\,\cite{Conway:1989fs}, rescaled according to the reanalysis described in Ref.\,\cite{Aicher:2010cb}, from which the dot-dashed (green) curve is drawn.
The dotted (red) curve is the result obtained using a Poincar\'e-covariant regularisation of a contact interaction, Eq.\,\eqref{xuCI}.
}
\end{figure}

In Fig.\,\ref{figxuvpi} we compare our result for the pion's valence-quark distribution with available experiment \cite{Conway:1989fs}.  In considering the data in Fig.\,\ref{figxuvpi}, it is important to recall that E615 \cite{Conway:1989fs} reported a PDF inferred via LO analysis in pQCD; and, as noted in Sec.\,\ref{secQDFs}, this yielded controversial behaviour on $x\simeq 1$, contradicting QCD-based expectations: producing $u^\pi(x) \sim (1-x)$ instead of $u^\pi(x)\sim(1-x)^2$.  A subsequent NLO reanalysis \cite{Aicher:2010cb}, which, crucially, also included soft-gluon resummation, indicated that the data are actually consistent with $u^\pi(x)\sim(1-x)^2$.  As emphasised by Ref.\,\cite{Wijesooriya:2005ir}, NLO evolution alone cannot expose that.  Thus, in Fig.\,\ref{figxuvpi} we plot the E615 data rescaled as follows ${\rm E615}_{\rm 2010} = \mathpzc{F}(x)\,{\rm E615}_{\rm 1989}$, where $\mathpzc{F}(x)$ is the $x$-dependent ratio of Fit-3 in Ref.\,\cite{Aicher:2010cb} to the E615 fit described in Table~VII of Ref.\,\cite{Sutton:1991ay}.  It is evident in Fig.\,\ref{figxuvpi} that the data and all QCD-based calculations agree on the behaviour of $u^\pi(x)$ within the valence-quark domain.  

In connection with the remarks made following Eq.\,\eqref{PDFQCD}, it is useful to report the pion valence-quark PDF obtained using a Poincar\'e-covariant regularisation of a momentum-independent (contact) quark-quark interaction, which is \cite{Roberts:2010rn} (chiral limit):
\begin{equation}
\label{CIPDF}
u_{CI}^\pi(x;\zeta_H) = \theta(x) \theta(1-x)\,.
\end{equation}
In the present application, this result is identical to that obtained using equivalent regularisations of the Nambu--Jona-Lasinio model (see, \emph{e.g}.\ Sec.\,VI.B.3 of Ref.\,\cite{Holt:2010vj} and citations therein).  Evolving this distribution as described in connection with Fig.\,\ref{figuKusratio}, one obtains
\begin{equation}
\label{xuCI}
x u_{CI}^\pi(x;\zeta_{5.2}) = 1.20 x^{0.73} (1-x)^{0.88},
\end{equation}
which is the dotted (red) curve in Fig.\,\ref{figxuvpi}.  Notably, evolving from a smaller initial scale, such as $Q_0^{\rm CI}=0.4\,$GeV, which is commonplace in applications of a contact interaction, has no material effect on the result in Eq.\,\eqref{xuCI}, \emph{viz}.\ $0.88 \to 1.1$, but both values are less-than 40\% of that required to be consistent with the modern reappraisal of E615 data \cite{Aicher:2010cb}, displayed in Fig.\,\ref{figxuvpi}.

Numerical simulations of lattice-regularised QCD (lQCD) typically report moments of hadron PDFs at a resolving scale $\zeta_2=2\,$GeV.  Importantly, owing to the loss of Poincar\'e-covariance, the most widely used lQCD algorithms only provide access to the lowest three nontrivial moments.   Such results are available for $u^\pi(x)$, \emph{e.g}.\ a contemporary simulation \cite{Brommel:2006zz}, using two dynamical fermion flavours, $m_\pi \gtrsim 0.34\,$GeV and nonperturbative renormalisation at $\zeta_2=2\,$GeV, produces the first row here:
\begin{equation}
\label{momentslQCD}
\begin{array}{l|lll}
    & \langle x \rangle_u^\pi & \langle x^2 \rangle_u^\pi & \langle x^3 \rangle_u^\pi\\\hline
\mbox{\cite{Brommel:2006zz}} & 0.27(1) & 0.13(1) & 0.074(10)\\
\mbox{\cite{Best:1997qp}} & 0.28(8) & 0.11(3) & 0.048(20)\\
\mbox{\cite{Detmold:2003tm}} & 0.24(2) & 0.09(3) & 0.053(15)\\
{\rm average} & 0.26(8) & 0.11(4) & 0.058(27)\\\hline
{\rm herein} & 0.26 & 0.11 & 0.052
\end{array}\,.
\end{equation}
The results in Ref.\,\cite{Brommel:2006zz} agree with those obtained in earlier estimates based on simulations of quenched lQCD \cite{Best:1997qp, Detmold:2003tm} and are consistent with the values obtained from our computed distribution, which are reported in the last row of Eq.\,\eqref{momentslQCD}.

Our predictions in Eq.\,\eqref{momentslQCD} are obtained via the LO-evolution of our result for $u^\pi(x;\zeta_H)$ to $\zeta_2$, which is satisfactorily interpolated by the form in Eq.\,\eqref{xqxfit} using the first row of coefficients below:
\begin{equation}
\label{xqxfitparamsz2}
\begin{array}{l|ccccc}
\zeta_{2}  & A &   \alpha  & \beta & \gamma  & \delta \\\hline
xu^\pi   &   \phantom{2}1.60   & 0.90   & 2.69   & 0 & 4.46 \\
xu^K     &   25.65 & 1.80   & 2.69   & 0.87 & 0\phantom{.46} \\
x{\bar s}^K     &   23.16 & 1.84   & 2.69   & 2.18 & 2.44
\end{array}\,.
\end{equation}
Plainly, all computations reported in Eq.\,\eqref{momentslQCD} agree that the valence-quarks carry only 50\% of the pion's light-front momentum at $\zeta_2$.  On the other hand, the contact interaction distribution in Eq.\,\eqref{CIPDF} yields the following values for the first three moments: $\{0.33,0.17,0.11\}$; and consequently predicts that two-thirds of the pion's momentum is carried by valence-quarks at this scale.

No lQCD results are yet available for moments of the kaon distributions; but our predictions for the first three moments of each distribution are:
\begin{equation}
\label{momentsK}
\begin{array}{l|lll}
q   & \langle x \rangle_q^K & \langle x^2 \rangle_q^K & \langle x^3 \rangle_q^K\\\hline
u & 0.28 & 0.11 & 0.048\\
\bar s  & 0.36 & 0.17 & 0.092
\end{array}\,.
\end{equation}
They are obtained from the LO-evolution of our results for $(u,{\bar s})^K(x;\zeta_H)$ to $\zeta_2$, which are satisfactorily interpolated by the form in Eq.\,\eqref{xqxfit} using the second and third rows of coefficients in Eq.\,\eqref{xqxfitparamsz2}. It is evident from these results that valence-quarks carry approximately two-thirds of the kaon's momentum at $\zeta_2$.

\section{Conclusions and possibilities}
\label{secEpilogue}
We employed a dressed-quark basis to analyse parton distribution functions of the pion and kaon.  The expressions that define these distributions [Eqs.\,\eqref{uvpix}, \eqref{valenceqK}] overcome weaknesses of the impulse approximation and ensure that, independent of model details, the dressed-quarks express a purely valence distribution, \emph{viz}.\ they always carry the entirety of a given meson's light-front momentum [Eqs.\,\eqref{symmuvpi}, \eqref{kaonmom}], and the valence-quark distribution behaves as $(1-x)^2$ on $x\simeq 1$ [Eq.\,\eqref{valencepower}].  Using algebraic formulae for the dressed-quark propagators and pion and kaon Bethe-Salpeter amplitudes, which express effects associated with both explicit and dynamical chiral symmetry breaking and produce the correct conformal-limit meson parton distribution amplitudes, we computed the valence dressed-quark PDFs for the pion and kaon [Fig.\,\ref{figPDFsQ0}].  The results demonstrate that it is the flavour-dependence of dynamical chiral symmetry breaking which modulates the strength of $SU(3)$-flavour symmetry breaking in meson PDFs.

We subsequently explained why, even at a typical hadronic scale, the valence dressed-quark structure of mesons as perceived in deep inelastic processes must be augmented by sea-quark and gluon contributions; and detailed a simple but realistic means of achieving this.  The corrections may be divided into two classes: [C1], which redistributes baryon-number and momentum into the dressed-quark sea; and [C2], which shifts momentum into the pion's dressed-gluon distribution.  Our analysis suggests that contributions within [C2] are most important at an hadronic scale, \emph{viz}.\ $\zeta_{\rm H} \approx 2\,\Lambda_{\rm QCD}$.

Working with this information, we built a simple algebraic model to express the principal impact of both classes of corrections on the pion and kaon, which, combined with the predictions of the dressed-quark basis, permitted a realistic comparison with existing experiment.  This enabled us to reveal essential features of these mesons' valence-quark distributions.
Namely, at a characteristic and reasonable hadronic scale, $\zeta_H$, valence dressed-quarks carry roughly two-thirds of the pion's light-front momentum, with the bulk of the remainder carried by glue.  In contrast, valence dressed-quarks carry approximately 95\% of the kaon's light-front momentum at $\zeta_H$, with the remainder lying in the gluon distribution.
This difference may be attributed to the fact that heavier quarks radiate soft gluons less readily than lighter quarks and momentum conservation communicates this effect to the kaon's $u$-quark.

Evolving our corrected distributions to a scale characteristic of meson-nucleon Drell-Yan experiments, we reproduced and explained extant data on the kaon-to-pion ratio of $u$-quark distributions [Fig.\,\ref{figuKusratio}] and the pion's $u$-quark distribution [Fig.\,\ref{figxuvpi}].
As a complement to these results, we also evolved the distributions to the resolving scale $\zeta_2=2\,$GeV, which is typically used in numerical simulations of lattice-regularised QCD.  Here, valence-quarks carry roughly one-half of the pion's light-front momentum but two-thirds of the kaon's momentum.

A valuable opportunity now presents itself.  Namely, it should be possible to employ the methods exploited in Refs.\,\cite{Chang:2013pq, Chang:2013nia, Shi:2015esa, Ding:2015rkn} and follow the procedures in Secs.\,\ref{secBuilding}, \ref{secDrawing} above so as to achieve a quantitatively reliable, QCD-connected unification of meson valence-quark distribution functions (PDF) with, \emph{inter alia}, their distribution amplitudes and elastic electromagnetic form factors.  Completing such a picture is crucial as hadron physics enters an era of new-generation experimental facilities capable of testing such an array of interrelated predictions.

\section*{Acknowledgments}
We are grateful for insightful comments and suggestions from I.\,C.~Clo\"et, P.~Hutauruk, C.~Mezrag, S.-X.~Qin and P.\,C.~Tandy.
Work supported by:
the U.S.\ Department of Energy, Office of Science, Office of Nuclear Physics, under contract no.~DE-AC02-06CH11357;
the Chinese Ministry of Education, under the \emph{International Distinguished Professor} programme;
the National Natural Science Foundation of China (grant nos.\ 11275097, 11275180, 11475085 and 11535005);
and the Fundamental Research Funds for the Central Universities Programme of China (grant no.\ WK2030040050).


\end{document}